# A Design Technique based on Equivalent Circuit and Coupler Theory for Broadband Linear to Circular Polarization Converters in Reflection or Transmission Mode

Gerardo Perez-Palomino, Juan. E. Page, Manuel Arrebola, *Member, IEEE,*
and José A. Encinar, *Fellow, IEEE*

*Abstract*—A new approach to designing FSS-based LP-CP converters is presented. It is based on the use of FSSs which exhibit dual diagonal symmetry, and a novel 4-port equivalent circuit able to describe the electrical behavior of the cells for the two linear incident polarizations at the same time. The equivalent circuit allows the use of standardized branch line coupler theory to design LP-CP converters comprising a cascade of an arbitrary number of layers, whose synthesis includes the phase and makes it possible to achieve prescribed electrical conditions systematically. A full design procedure has been developed using the new approach and several designs in both transmission and reflection modes are presented and evaluated. It has been proven that single layer reflective converters exhibit large bandwidths as the two reflected field components are in quadrature independently of the frequency. One of these devices was designed and showed an AR<0.2 dB within the band from 21.5 to 28.5 GHz. The reflective LP-CP converter designed was also manufactured and tested, and the measurements were used to validate the design procedure.

## I. Introduction

LINEAR to circular polarization (LP-CP) converters are devices widely used in optical and microwave systems in applications that need circular polarization (CP) to be taken advantage of. Their architectures depend on the type of wave to be converted, which can be either guided [1]-[2] or propagated in free space. In the latter case, the most common cell architecture used to obtain LP-CP converters for both modes of operation (transmission or reflection) is based on stacked planar Frequency Selective Surfaces (FSSs), since they can be easily manufactured and integrated. Thus, efforts have been made for a long time to study them [3]-[19].

The operating principle of all these works assumes the approach that considers an incident electric field, linearly polarized tilted 45º with respect to one of the two axes that define the period of the cell (see Fig. 1a), dividing the field into two equal orthogonal components. In this case, the cascade of FSSs must be designed to introduce a relative phase-shift of 90º between the two transmitted/reflected field components as well as a unitary module for the incident field.

The work described in [3] shows one of the first LP-CP converters based on an FSS architecture, which was designed using a cascade of rectangular FSSs to synthesize a combination of both capacitive and inductive impedance sheets. In [4]-[7], meander-line shapes were researched, which exhibit good electrical behavior in terms of the most relevant features for this type of device (Axial Ratio, bandwidth, reflection levels, or range of angles of incidence), as well as simplicity in the design procedure because the meanders are capacitive for one polarization and inductive for the orthogonal one. Other element shapes based on split-rings [8]-[9], crosses [10]-[11], or dipoles fed by a slotted planar waveguide [12] were also proposed to improve the electrical performance and the integration of the feeder.

The concept of metasurface has been also used in recent years to develop LP-CP converters, which involve cells made up of sub-wavelength structures which are electrically thin, thus reducing the profile of the device [13]-[19]. Several metasurface-based designs have been reported to operate in reflection [13]-[14], transmission [15]-[17] or hybrid mode [18], some of which integrate the feed system in the stacking [17] or include beam focusing and gain enhancement [13].

The strategies used in designing LP-CP converters in these works are varied. In [3]-[8] the design procedure is based on the use of equivalent circuits combined with electromagnetic (EM) simulators. The circuits are used to obtain the values of susceptances that fulfill the design conditions for the two polarizations efficiently, whereas the EM simulators are only used to find the correspondence between the dimensions of the cell and the circuital parameters. In these cases, if the cell

Manuscript received July 26, 2017. This work was supported in part by the Community of Madrid, under S2013/ICE-3000 (SPADERadar-CM) project, by the Spanish Ministry of Economy, Industry and Competitiveness, under the project TEC2016-75103-C2-1-R, and by the European Space Agency, under project 4000117113.
G. Perez-Palomino, J. E. Page, and J.A. Encinar are with the Department of Electromagnetism and Circuit Theory, Universidad Politécnica de Madrid, E-28040, Madrid, Spain (e-mail: gperez@etc.upm.es).
M. Arrebola is with Signal Theory and Communications Group, Department of Electrical Engineering, Universidad de Oviedo, Campus Universitario 33204–Gijon, Spain (e-mails: arrebola@uniovi.es).

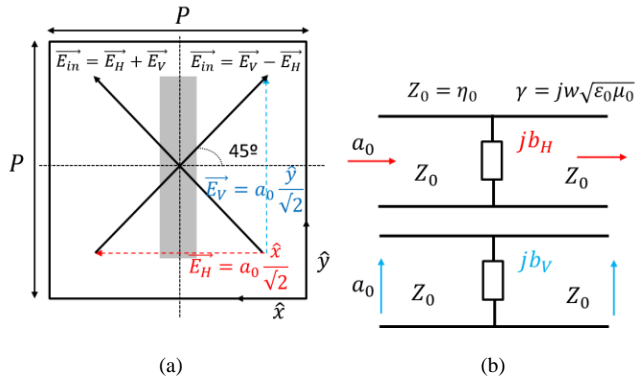

Fig. 1. (a) Architecture of an FSS cell illuminated by a linearly polarized electric field tilted 45º, (b) Equivalent circuit of the cell at normal incidence for vertical (V) and horizontal (H) polarization.

architecture is suitable, the topology of the circuit models can allow the use of standard synthesis to calculate the required susceptances [3]. On the other hand, if the circuit model does not allow standard synthesis, optimization processes must be applied to the equivalent circuits [15], [18]. However, even in this case the efficiency is quite acceptable when compared with the designs that use an EM simulator in the full design process, which is usually done when an equivalent circuit cannot describe the electrical behavior of the cells [9], [17].

In the cases in which the circuits can be used, the approach that considers a linear incident field tilted 45º makes use of two independent equivalent circuits, one for each polarization (Fig. 1b). Thus, the design in the case of the transmission mode involves the synthesis of two different filters that ensure the same transmission ripple for the two linear polarizations, while the required phase-shift of 90º is also achieved within the band. However, although the cell topology makes it possible to use standard synthesis to design both filters [20], it is only limited to the amplitude response, so that the phase restrictions are not included in the synthesis. The strategy commonly used to reach the phase-shift requirement within a certain bandwidth consists of designing two filters centered at a different frequency to each other, so that the separation between their central frequencies produces a phase-shift of 90º [15]. However, two filters designed at different central frequencies exhibiting the same order and ripple require the use of two different separations, which is not physically achievable. Therefore, this strategy involves a different ripple for each filter, which impairs the performance even at the central frequency, $f_0$, unless further optimizations allow the problem to be overcome. Thus, the fact that phase restriction is not considered in the circuital synthesis could give rise to impairments in the electrical performance and/or a more complex design procedure that includes additional iterations and optimizations.

In this paper, a new approach to designing FSS-based LP-CP converters is presented. The approach is based on the use of FSSs exhibiting dual diagonal symmetry and a novel equivalent circuit that is introduced for first time and is able to describe the electrical behavior of the cells. The properties of the diagonal symmetries make it possible to describe the FSS

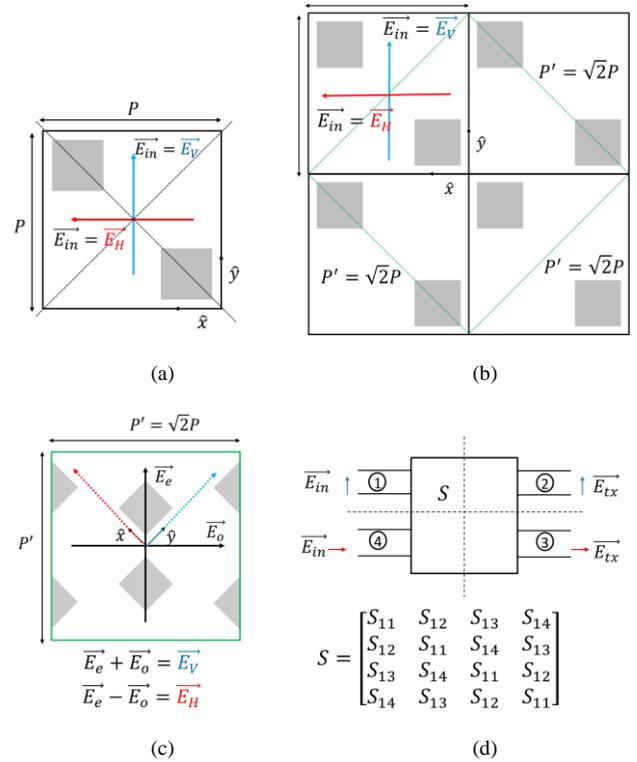

Fig. 2. (a) Architecture of FSS cell with dual diagonal symmetry illuminated by a linearly polarized electric field tilted 0º, (b) Extended view of the infinite FSS involving a group of 4 cells where the equivalent cell can be observed, (c) Equivalent cell obtained by rotating 45º the original cell, and (d) S matrix of a dual symmetric 4 ports network

cell for the two linear incident polarizations (at 0º and 90º, see Fig. 2a) at the same time using a single 4-port circuit, which in turn allows the use of standardized branch line coupler theory [20] in the design of LP-CP converters made up of a cascade of an arbitrary number of layers in both transmission and reflection modes of operation. The standardized process could provide some advantages in the design such as simplicity and better performance, since the phase can be included in the synthesis systematically. Furthermore, the use of a standard synthesis is a powerful tool in establishing the electrical limits of the device, since it makes it possible to predict the minimum number of layers required to reach a certain reflection level, bandwidth and axial ratio (AR).

A full design procedure has been developed and validated using this new approach. Several designs in both reflection and transmission are presented making use of the same FSS shape, whose electrical limits are studied. Note that the cells showing dual diagonal symmetry also make it possible to use FSS-scatters that are usually difficult to imagine using the other approach (mainly because of their structural complexity), which could provide better performance than other shapes reported in the literature. One of these designs, which has been manufactured, measured and used to validate the full design procedure, is presented here.

## II. THEORETICAL BASIS AND CIRCUIT MODEL

Let us consider a square period single layer FSS, whose metallization thicknesses are negligible, placed in a perfect vacuum. It is well known that if the FSS cell exhibits two symmetries in both horizontal (H) and vertical (V) planes (as shown in Fig. 1a), the electrical response of the cell at normal incidence and H-polarization does not generate a vertical component, and vice versa. Thus, both polarizations are decoupled and can be described independently. Since the FSS cell behaves like a transmission line made up of two parallel plates, the equivalent circuit for each polarization is known and can be represented as a parallel admittance (a susceptance in the absence of losses) whose value depends on the metallization (see Fig. 1b). However, this behavior is not produced in the cell shown in Fig. 2a due to the absence of symmetry in the main planes. In this case, the cell will generate both field components for a (V) or (H) incident polarization, so that the electrical response must be described with a 4-port circuit. At present, the elements and the architecture of the 4-port network that allow an arbitrary FSS cell to be described have not yet been found because of the complexity of the problem. These elements have been obtained here for first time if the FSS exhibits dual diagonal symmetry (as shown in Fig. 2a).

To find the equivalent circuit in this particular case, let us consider the equivalent cell obtained when the FSS shown in Fig. 2a is rotated 45° (see Fig. 2b). It is clear that the equivalent cell (represented in Fig. 2c) shows a new period ($\sqrt{2}P$) and can be excited by two modes: odd ($\vec{E_o}$) and even ($\vec{E_e}$), according to the boundary conditions of the problem. Both modes form an orthogonal basis that makes it possible to describe any arbitrary excitation, and particularly the two original (V) and (H) modes as a sum and difference, respectively.

If the even and odd modes are considered simultaneously, the resulting network would be that represented in Fig. 2d, which has two symmetries and four independent scattering parameters. To characterize the 4-port network, the conventional odd and even decomposition is applied, so that each of the two excitations corresponds to a symmetric quadripole:

$$S_e = \begin{pmatrix} S_{11e} & S_{12e} \\ S_{12e} & S_{11e} \end{pmatrix}, \quad S_o = \begin{pmatrix} S_{11o} & S_{12o} \\ S_{12o} & S_{11o} \end{pmatrix} \quad (1)$$

whose parameters can be associated with a parallel admittance, $Y_e$ and $Y_o$, so that:

$$Y_e = jb_e = \frac{-2S_{11e}}{Z_0 S_{12e}} \quad Y_o = jb_o = \frac{-2S_{11o}}{Z_0 S_{12o}} \quad (2)$$

Then the **S** matrix (referred to $Z_0$) can be written as:

$$S_{11} = \frac{S_{11e}+S_{11o}}{2} \quad S_{12} = \frac{S_{12e}+S_{12o}}{2} \quad (3)$$

$$S_{13} = \frac{S_{12e}-S_{12o}}{2}, \quad S_{14} = \frac{S_{11e}-S_{11o}}{2}$$

where

$$S_{11} = \frac{-1}{2}\left(\frac{Z_0 Y_e}{2+Z_0 Y_e} + \frac{Z_0 Y_o}{2+Z_0 Y_o}\right) \quad (4)$$

$$S_{12} = \frac{1}{2+Z_0 Y_e} + \frac{1}{2+Z_0 Y_o} \quad (5)$$

$$S_{13} = \frac{1}{2+Z_0 Y_e} - \frac{1}{2+Z_0 Y_o} \quad (6)$$

$$S_{14} = \frac{-1}{2}\left(\frac{Z_0 Y_e}{2+Z_0 Y_e} - \frac{Z_0 Y_o}{2+Z_0 Y_o}\right) \quad (7)$$

Since the **S** parameters are related to a 4-port network that must be reduced to parallel admittances when a diagonal excitation is applied, the equivalent circuit must exhibit the architecture shown in Fig. 3a, whose connection quadripole in transmission parameters, $T_c$, is:

$$T_c = \begin{pmatrix} A & B \\ C & D \end{pmatrix} = \begin{pmatrix} \frac{Y_o+Y_e}{Y_o-Y_e} & \frac{2}{Y_o-Y_e} \\ \frac{2Y_o Y_e}{Y_o-Y_e} & \frac{Y_o+Y_e}{Y_o-Y_e} \end{pmatrix} \quad (8)$$

and whose total transmission matrix, **T**, is (**T** is selected to analyze a stacked FSS architecture more easily):

$$\begin{pmatrix} V_1 \\ I_1 \\ V_4 \\ I_4 \end{pmatrix} = \underbrace{\begin{pmatrix} 1 & 0 & 0 & 0 \\ \frac{Y_e+Y_o}{2} & 1 & \frac{Y_e-Y_o}{2} & 0 \\ 0 & 0 & 1 & 0 \\ \frac{Y_e-Y_o}{2} & 0 & \frac{Y_e+Y_o}{2} & 1 \end{pmatrix}}_{T} \cdot \begin{pmatrix} V_2 \\ I_2 \\ V_3 \\ I_3 \end{pmatrix} \quad (9)$$

The equivalent circuit may be generalized to consider the metallization of non-negligible thicknesses, and allows the presence of dielectric layers such as supporting materials or spacers. The latter must be introduced into the circuit by adding transmission lines whose impedances and lengths are appropriate.

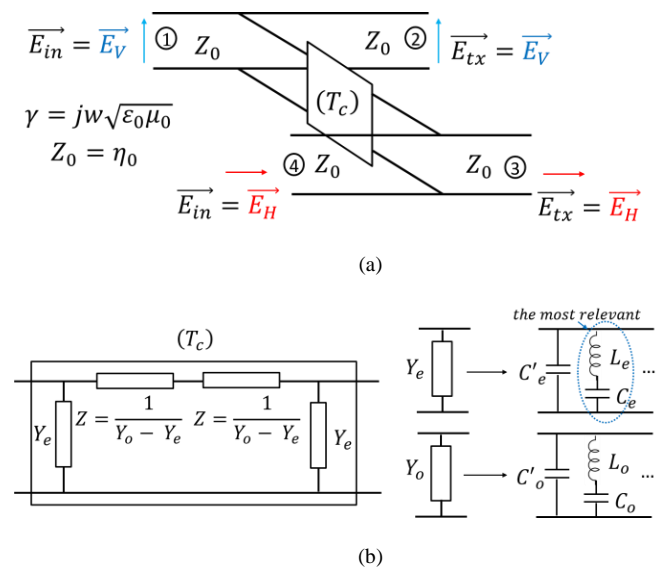

Fig. 3. (a) Equivalent circuit of the cell shown in Fig. 2a, (b) Pi-representation of the connection quadripole and Foster representation of the even and odd admittances

As usual, the use of the equivalent circuit requires the prior calculation of the admittances, $Y_e$ and $Y_o$, using an EM simulator. These parameters can also be expressed as a Foster expansion (as shown in Fig. 3b) to obtain an equivalent circuit of lumped elements within the entire band of viability of the circuit [21], which covers the band within which the first harmonic is the only propagating mode in the cell for both polarizations. The first high-order harmonics are the $TE_{20}$ and $TM_{02}$ modes of the parallel-plate waveguide because of the two symmetries of the rotated cell. Both modes show a cut-off frequency given by $f_c = c_0/P'$, where $P' = \sqrt{2} \cdot P$ is the period of the equivalent cell (Fig. 2c). Therefore, $f_c$ is the maximum frequency of validity of the equivalent circuit.

It should be noted that the proposed circuit also assumes that the evanescent harmonics are sufficiently attenuated between two stacked FSSs (12-15 dB), as the interactions of these modes with the following discontinuity are not considered when the even and odd admittances are calculated. This phenomenon, which depends on both period and frequency, limits the separation allowed between two stacked FSSs to make the equivalent circuit sufficiently accurate. Thus, separations greater than $\lambda/8$ would be desirable ($\lambda$ being the wavelength), although in practice the thicknesses could be even thinner to obtain suitable circuital pre-designs.

## III. DESIGN OF REFLECTIVE LP-CP CONVERTERS

In the case of reflective LC-CP converters, equations (4)-(7) indicate that a single layer FSS is not able to convert the polarization if the transmission is null, except in the case in which the output ports are charged with a pure reactance ($Y_L=jb_L$). Therefore, the FSS architecture involves placing a short circuit (conducting surface) at a certain distance from the scatter. Thus, the simplest cell necessary to design the converter, which is made up of two stacked dielectric layers, is shown in Fig. 4a. The first layer is a supporting substrate upon which the FSS is printed, and the second one could be a mechanical substrate or vacuum. The equivalent circuit is then represented as in Fig. 4b, which is a two-port circuit whose main parameters can be easily calculated as:

$$\rho_{11} = \frac{\frac{1}{Z_0^2} - Y_e Y_L - Y_o Y_L - Y_e Y_o - Y_L Y_L}{\left[\frac{1}{Z_0} + Y_e + Y_L\right]\left[\frac{1}{Z_0} + Y_o + Y_L\right]} \quad (10)$$

$$\tau_{12} = \frac{Y_o - Y_e}{Z_0\left[\frac{1}{Z_0} + Y_e + Y_L\right]\left[\frac{1}{Z_0} + Y_o + Y_L\right]} \quad (11)$$

From (10) and (11), it is easy to deduce that both ports are in quadrature at an arbitrary frequency ($\Delta\varphi = \arg(\rho_{11}) - \arg(\tau_{12}) = \pm\pi/2$). Although the reflected polarization is usually elliptical for a certain incident linear polarization, if:

$$\left|\frac{\rho_{11}}{\tau_{12}}\right| = \left|Z_0 \frac{\frac{1}{Z_0^2} - Y_e Y_L - Y_o Y_L - Y_e Y_o}{Y_o - Y_e}\right| = 1 \quad (12)$$

then the reflected polarization is circular, which will be right or left handed for an incident vertical polarization if $\varphi = -\pi/2$ and $\varphi = \pi/2$, respectively.

In the particular case in which $Z_s = Z_1 = Z_0$, the load admittance at $f_0$ is $Y_L = -jY_0\cot(k_0 d)$, and (12) leads to four different analytical solutions for the length of the line, $d$:

$$d = \frac{1}{k_0}\cot^{-1}(-Z_0 b_L) \quad (13)$$

where

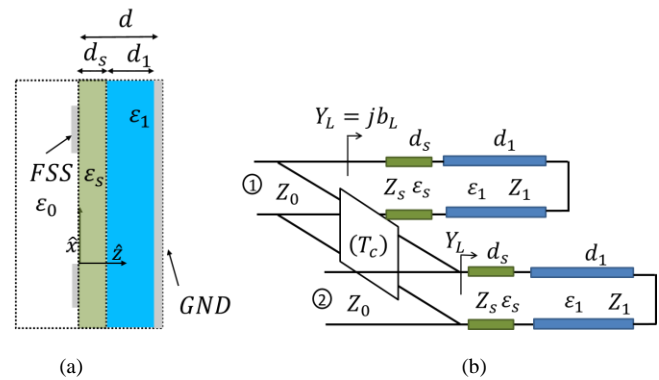

Fig. 4. General architecture of a single layer reflective LP-CP converter made up of FSS with dual diagonal symmetry (Fig. 2a), (a) side view, (b) Equivalent circuit

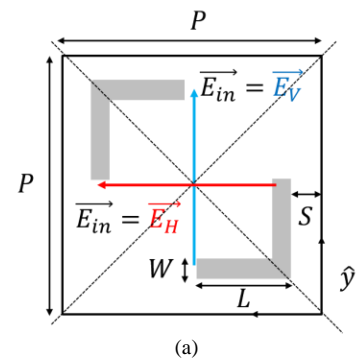

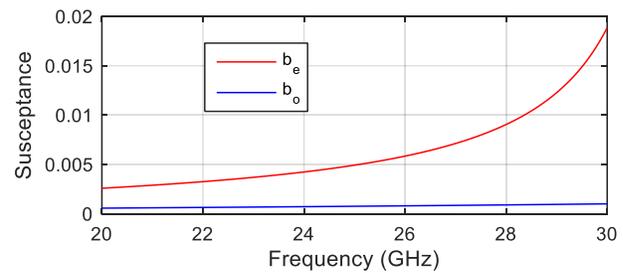

Fig. 5. FSS shape used to make the designs in this paper, (a) top view (b) Even and odd susceptances in the vacuum

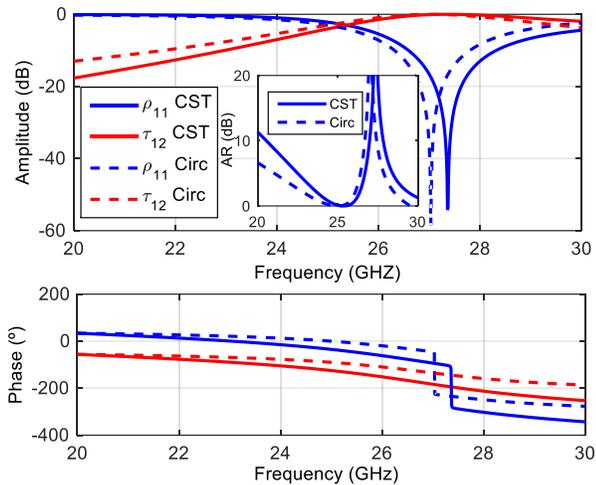

Fig. 6. Circuital (Circ) and electromagnetic (CST) responses of the design of a single layer LP-CP converter when considering: $P = 5\ mm$, $L = 2.5\ mm$, $W = 0.5\ mm$, $S = 0.075\ mm$, $d_1 = 0.828\ mm$, $d_s = 0\ mm$, $\varepsilon_1 = 1$, $\varepsilon_s = 3.38$.

$$b_L = -\frac{b_e + b_o}{2} \pm \sqrt{\left(\frac{b_e + b_o}{2}\right)^2 - \frac{1}{Z_0^2} - b_e b_o} \pm \frac{b_e - b_o}{Z_0} \quad (14)$$

Note that the values of the lengths are obtained at the central frequency, $f_0$, for a certain fixed geometry of the FSS-scatter, so that if there is no physical solution to (13), the scatter must be changed.

The use of the equivalent circuit shown in Fig. 4b also allows for designing a multi-layer architecture charged with a ground plane. The design in this case will be considered in the next section, which focuses on the design of LP-CP converters in transmission mode.

*A. Example 1:*

As a particular design of a reflective LP-CP converter, a single layer design has been made in accordance with the stacking shown in Fig. 4a, where the material permittivities are: $\varepsilon_s = 3.38$ and $\varepsilon_1 = \varepsilon_0 = 1$. The FSS-scatter used in the design is shown in Fig. 5a, which was selected because it provides three design parameters: length (L), width (W) and separation (S). This particular scatter was introduced in [22] to design metamaterial absorbers.

The design will be made to operate at the central frequency, $f_0$=25 GHz, and the selected upper limit of the bandwidth of interest was 30 GHz. The period P=5 mm was chosen to be half of a wavelength at the upper limit, so that the first harmonic is propagated at much higher frequencies (fc=42.42 GHz). The values L=2.5 mm, W=0.5 mm, were selected as the initial parameters, and "S" was small to obtain high capacitance values (S=0.075 mm). In this case, the values of the even and odd susceptances for several frequencies when the reference impedance is $Z_0 = \eta_0$ are represented in Fig. 5b, which can be associated with a series resonator and therefore with capacitances and inductances (Ce, Le and Co, Lo). In a first approach, the distances to the ground plane are calculated by considering that $d_s \ll d_1$, so that (14) provides two implementable distances: $d = 0.828\ mm$ and $d = 3\ mm$. The selection of the appropriate distance should take into account the limitations of the equivalent circuit in terms of the effects of high order harmonics. Thus, if the separation is too small, the attenuation of the first harmonic might not be high enough to make its effect negligible even when it is not propagating, as previously described in section II. Therefore, an increase in the length of the line of $\lambda/2$ at the central frequency should be assumed in this case.

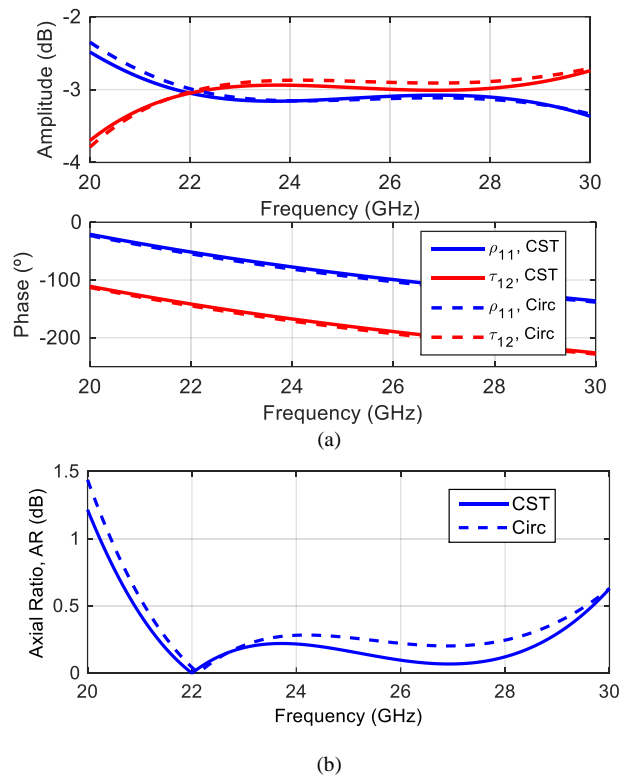

Fig. 7. Circuital (Circ) and electromagnetic (CST) responses (normal incidence) of the design of a single layer LP-CP converter when considering: $P = 5\ mm$, $L = 2.1\ mm$, $W = 0.55\ mm$, $S = 0.075\ mm$, $d_1 = 1.5\ mm$, $d_s = 0.5\ mm$, $\varepsilon_1 = 1$, $\varepsilon_s = 3.38$, $tan\delta_s = 0.003$. (a) Amplitude and phase of the reflection ($\rho_{11}$) and conversion ($\tau_{12}$) coefficients, (b) AR

Fig. 6 shows the amplitude and phase of $\rho_{11}$ and $\tau_{12}$ (the design that considers d=0.828 mm) calculated using two different tools: the equivalent circuit and the electromagnetic simulator (CST [23]); the AR is also represented in both cases. As can be seen, the equivalent circuit is able to describe the electromagnetic behavior of the cell with relatively good accuracy, although some discrepancies are observed mainly due to the reflections of the first high-order harmonics on the ground plane (d=0.828 is too thin to appropriately attenuate these harmonics). However, the accuracy is enough to carry out the design with a guarantee of success, so it can be used in the first steps of the design procedure to obtain a good starting point that can be used by the electromagnetic simulator. Also note that the phase difference between the reflection and transmission parameters is exactly 90º at any frequency, as demonstrated in (10) and (11). Thus, a broadband design only means that the amplitudes of both parameters are taken into account.

The curves shown in Fig. 6 are associated to the design when the dimensions are: L=2.5 mm, and W=0.5 mm. However, it could be possible to find other pairs of values of length and width that provide a design exhibiting better performance in terms of bandwidth. In fact, a broadband design implies finding the appropriate pair to maximize the bandwidth. To achieve this goal, the procedure used here was: (1) to simulate 25 pairs of values of the length "L" and width "W" electromagnetically (S=0.075 mm was fixed), (2) to calculate the lumped elements of the even and odd susceptances of each pair (capacitances and inductances: Ce, Le, Co, Lo), since they provide an equivalent circuit whose components are independent of the frequency, (3) to interpolate the values of the even/odd capacitances and inductances from the 25 samples with the aim of generating a data basis that relates the circuital parameters and the physical dimensions, (4) to calculate the corresponding values of "d" (analytical), and (5) to find the parameters that maximize the bandwidth using the equivalent circuit. Note that the searching process is carried out at the circuital level, so that it is computationally efficient. Once the values of ("L", "W" and "d") are found from the circuital analysis, they were used as a starting point in an electromagnetic simulator to optimize the structure.

After applying this circuital procedure, the values of the even and odd capacitances and inductances were: $C_e$=4.7 pF, $C_o$=2.23 pF, $L_e$= 6.22 pH, $L_o$=4.39 pH, and the related dimensions: $L = 2.1\ mm$, $W = 0.55\ mm$, $S = 0.075\ mm$, $d_1 = 1.54\ mm, d_s = 0.5\ mm$. Since the effect of the supporting dielectric is not negligible in this case, it was included in the circuital model. The equivalent circuit represented in Fig. 4b is suitable, although it has to be taken into account that the values of the odd and even susceptances must be calculated to include the effect of the change of dielectric in the discontinuity. Once the first set of dimensions was obtained, a slight optimization was carried out with the aim of obtaining a design based on commercial materials. Thus, a supporting material of $d_s = 0.5\ mm$, $tan\delta_s = 0.003$ and $d_1 = 1.5\ mm$ was finally considered. The final dimensions after the optimization do not vary significantly with respect to the first set of dimensions. Fig. 7a shows the amplitude and phase of the reflection and polarization conversion coefficients, $\rho_{11}$ and $\tau_{12}$, as a function of the frequency at normal incidence, calculated with both the equivalent circuit and the electromagnetic simulator CST. It can be appreciated that a linear vertical incident polarization (port 1) is reflected into both the vertical (port 1, $\rho_{11}$) and the horizontal (port 2, $\tau_{12}$) components with the same proportion (around 3 dB) within a large bandwidth. Since the phase-shift between the V and H reflected components is exactly 90º, the reflected wave exhibits circular polarization (left handed) with the AR shown in Fig. 7b. It is obvious that the symmetry of the structure makes that a right handed circularly polarized wave with the same performance is reflected when a linear horizontal field impinges on the converter. The AR obtained is better than 0.2 dB within a large bandwidth from 21.5 to 28.5 GHz (29 %).

As regards the behavior of the converter designed at an oblique incidence, Fig. 8 represents the AR at different angles of incidence. As can be seen, the performance becomes worse as the angle increases from the normal. Note that the equivalent circuit is only valid at normal incidence, so that the aforementioned design procedure is not suitable for obtaining an efficient design at an oblique incidence. However, as usual in the literature, the design at normal incidence can be used as the basis for making particular designs at other angles of incidence with the aid of an electromagnetic simulator. As an example, Fig. 8 also shows the AR of the structure particularly optimized at 20º (TE), which improves the AR by 1 dB.

It is also possible to design single layer reflective converters using other dielectric materials, which are especially suitable for manufacture provided that the resulting thickness is commercially available. Fig. 8 also shows the AR of a reflective single layer converter designed to operate at an angle of incidence of 20º (TE) using ($\varepsilon_1 = \varepsilon_s = 3.38$). In this case, the period was considered as a design parameter.

The results show that the FSS used in Fig. 5 (or its equivalent when rotated 45º) exhibits a better electrical performance in terms of bandwidth than those reported in the literature for single layer reflective converters (single-resonant or multi-resonant elements [13], [24]-[25]). The results are comparable or even better than those presented in [14], which also achieve a very large bandwidth although using a more complex cell architecture.

IV. DESIGN OF LP-CP CONVERTERS IN TRANSMISSION

The equivalent circuit shown in Fig. 3a can be used to design LC-CP converters that operate in transmission mode. In this case, if the four ports of the circuit are charged with the reference admittance, $Y_0$, (4)-(7) imply the impossibility of achieving full transmission for a linear incident polarization (V or H). Therefore, an LP-CP converter cannot be designed by only using a single-layer FSS, the use of at least two layers being necessary.

The architecture made up two FSS layers (with dual diagonal symmetry) separated by a dielectric of arbitrary permittivity, $\varepsilon_1$, and thickness, d, is represented in Fig. 9a. In this case, the equivalent circuit is shown in Fig. 9b, which shows a similarity with a branch line coupler. Thus, if a separation of d=λ/4 is considered (λ is the wavelength at the design frequency), the matching condition that achieves the same amplitudes in ports 2 and 3, would involve a branch line design of 3 dB and the use of the permittivity $\varepsilon_1 = 2$. In accordance with branch line coupler theory, the even and odd admittances must fulfill ($Z_{ref}=Z_0$):

$$Y_e = -Y_o = j/Z_{ref} \qquad (15)$$

so that if the FSSs are designed to comply with this equation, ports 2 and 3 of the equivalent circuit would exhibit the same module and a phase-shift of 90°, and a circularly-polarized wave would be transmitted with no reflections (ports 1 and 4).

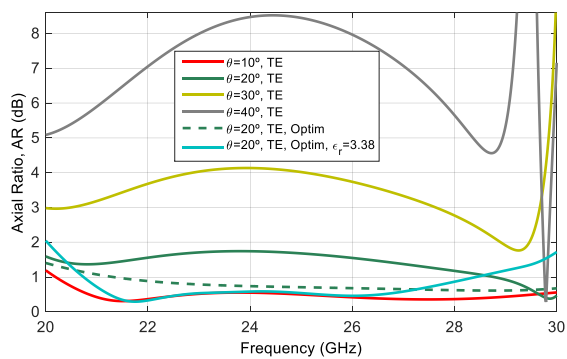

Fig. 8. Axial Ratio of the design of Fig. 7 for several angles of incidence that includes the response of a particular optimized design at 20º whose dimensions are (Optim): $P = 5\ mm, L = 2.1\ mm, W = 0.53\ mm, S = 0.075\ mm,\ d_1 = 1.64\ mm, d_s = 0.5\ mm, \varepsilon_1 = 1,\ \varepsilon_s = 3.38$ and (Optim $\varepsilon_r = 3.38$): $P = 4.5\ mm, L = 1.88\ mm, W = 0.53\ mm, S = 0.075\ mm, d_1 = 1.56\ mm, d_s = 0\ mm, \varepsilon_1 = \varepsilon_s = 3.38$

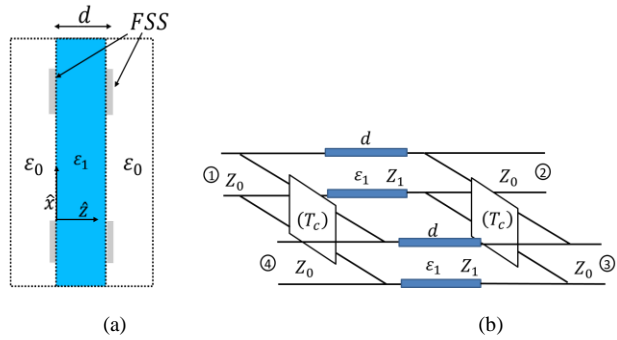

Fig. 9. General architecture of a dual layer LP-CP converter in transmission made up of FSSs with dual diagonal symmetry (Fig. 2a), (a) side view, (b) Equivalent circuit

In the general case of a multi-layer structure made up of "N+1" stacked FSSs separated by "N" dielectric spacers, the synthesis can be generalized to the design of a 3 dB branch line coupler of "N" sections, as shown in Fig. 10. The synthesis is standard, and allows different design possibilities and frequency behaviors: periodic, maximally flat, Chebyshev, etc [20], which requires the use of λ/4 transmission lines. The periodic synthesis is especially suitable, because it provides the same impedance as the reference for every spacer, so that they can be air layers or low permittivity dielectrics. If the multi-layered FSSs architecture is loaded with a ground plane (multi-layer reflective LP-CP converter), the architecture also allows a standardized synthesis. Each type of synthesis provides the admittance values of both the spacers and branches, and therefore the values of the even and odd susceptances to be achieved for each FSS. Thus, the synthesis is reduced to ensure that the FSS of each layer reaches (15) independently for the appropriate reference impedance.

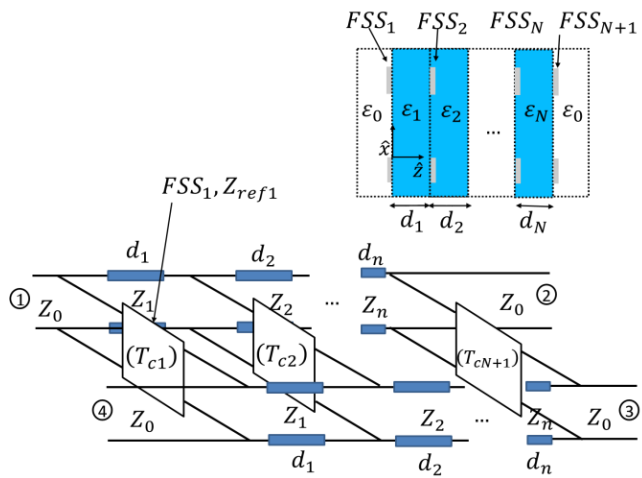

Fig. 10. Equivalent circuit of a LP-CP converter working in transmission mode made up of "N+1" FSS layers with dual diagonal symmetry (Fig. 2a)

It is clear that the even and odd admittances of an FSS cannot fulfill (15) at every frequency, since the FSS usually behaves as a series resonator (Fig. 5b). However, it is expected that this condition can be accomplished at least at the central frequency. Therefore, the resulting designs will usually present lower bandwidths than those using direct transmission lines.

Note that the synthesis of 3 dB branch-line couplers of N sections is able to guarantee that both transmitted linear polarizations exhibit the same amplitude within a prescribed bandwidth, ripple and reflection level. As regards the achievement of a phase difference of 90° between both field components, the synthesis also ensures that the phase-shift of 90° tends to be achieved within the prescribed bandwidth under a certain phase ripple (90° is achieved exactly at the central frequency). Thus, the bandwidth of the LP-CP converter will increase inherently for a given FSS shape as the number of sections increases. Since an FSS usually behaves as a series resonator, it could be possible to relate the prescribed parameters of the synthesis (bandwidth, ripple of the coupling, reflection levels and number of layers) with the AR, and therefore predict the full electrical performance of the device if the FSS used reaches the appropriate circuital values.

Once the synthesis is carried out and the correspondence between the circuital and geometrical parameters is calculated, the design is relatively easy. Since several combinations of the dimensional parameters could provide a solution at the central frequency, the equivalent circuit can be used to find the appropriate parameters to maximize the bandwidth within the limits of the FSS used. In this case, the computation of the response of the device is quite simple if the four-port transmission matrix, **T**, is used to define each FSS layer and spacer, as the behavior of the stacked architecture can be easily obtained by multiplying the transmission matrices of each of the components of the multi-layered structure.

It should be noted that this method cannot be generalized to consider a rectangular lattice, as shown in [19], [26]. In this case, the absence of dual diagonal symmetry makes that the equivalent four-port circuit shows asymmetric impedances in both the ports and the connection quadripole, thus not making it possible to use the standard branch line coupler theory.

The use of a standardized synthesis provides beneficial conditions for the designer, since it provides: (1) a systematic design procedure that includes the phase and allows for the prescribed electrical conditions to be achieved, (2) theoretical limits to the fractional bandwidth to be reached with a certain number of layers (the maximum limit of the structure that does not depend on the geometry of the FSS) and/or (3) the electrical limits of the FSS-scatter selected. However, there are some limitations that make the method does not take advantage of all possible degrees of freedom of a transmissive cell comprising N arbitrary FSSs, which could limit its electrical performance: (a) the standard synthesis requires fixed values of thickness and impedance for every dielectric layer (multiple of a quarter of a wavelength), so that the designed converters show a relatively high profile, (b) the equivalent circuit is suitable if the cell exhibits: square period, dual diagonal symmetry, single mode operation and enough attenuation of the evanescent modes, as discussed in section II, (c) the cell must operate in far-field conditions at normal incidence. Note that although the equivalent circuit is not suitable at oblique incidence, the designs at normal incidence are useful as an optimum base to use electromagnetic simulators and optimization processes, as demonstrated in section III.

If these limitations can be accepted, a standard and systematic design procedure is possible for an arbitrary number of layers, which could be especially beneficial in developing LP-CP converters comprising a large number of layers needed to fulfill stringent band requirements for both AR and reflection. Otherwise, other design strategies should be considered [7]-[19], although at the expense of non-standardized methods that in the best scenario require optimizations to be applied to the circuit. If (b) and (c) are assumed, the equivalent circuit presented in this paper could be still used to obtain hybrid design strategies that imply the optimization at the circuital level to reduce the thickness of the dielectric layers and the profile of the device.

### A. Example 2

A two-layer LP-CP converter in transmission has been designed to operate at the central frequency $f_0$=25 GHz, which is made up of the FSS shape used in section III.A (Fig. 5). As previously mentioned, the permittivity of the dielectric spacer must be $\varepsilon_1 = 2$, and the thickness $d = \lambda_1/4 = 2.121$ mm, although in the design the thickness has been increased $\lambda/2$ to achieve a good accurate equivalent circuit ($d = 3\lambda_1/4$ =6.364 mm). The even and odd susceptances for a set of dimensional values that fulfills (15) at 25 GHz is represented in Fig. 11a ($L = 3.4\ mm, W = 0.1\ mm,\ S = 0.075\ mm, P = 5\ mm$).
Note that condition (15) cannot be fulfilled because the odd susceptance is positive within the bandwidth within which the even and odd susceptances have different sign to each other, contrary to what is necessary. However, this problem can be easily solved by considering lengths of $3\lambda_0/4$ instead of $\lambda_0/4$ for the two branches (the length of the spacer is not affected). The values of the even and odd capacitances and inductances at each FSS layer are: $C_{e1}$= $C_{e2}$=16.75 pF, $L_{e1}$= $L_{e2}$=8.16 pH, $C_{o1}$= $C_{o2}$=2.82 pF and $L_{o1}$= $L_{o2}$=8.95 pH. After a small optimization, the dimensions of the final design were: L=3.25 mm, W=0.1 mm, S=0.075 mm and $d = 6.41\ mm$.

Fig. 11b and Fig. 11c show the reflections and the AR calculated with CST of the final design at normal incidence and linear vertical incident polarization. As can be seen, the device exhibits a small bandwidth, as expected from a 3-dB

TABLE I
CAPACITANCES AND INDUCTANCES TO PERFORM THE THREE LAYER LP-CP CONVERTER IN TRANSMISSION WITH PERIODICAL SYNTHESIS AT 25 GHZ

| FSS Layer | Even | Odd |
|---|---|---|
| 1 | C=19 pF, L=17.5 pH | C=1.2 pF, L=18.5 pH |
| 2 | C=40.7 pF, L=10 pH | C=2 pF, L=11.5 pH |
| 3 | C=19 pF, L=17.5 pH | C=1.2 pF, L=18.5 pH |

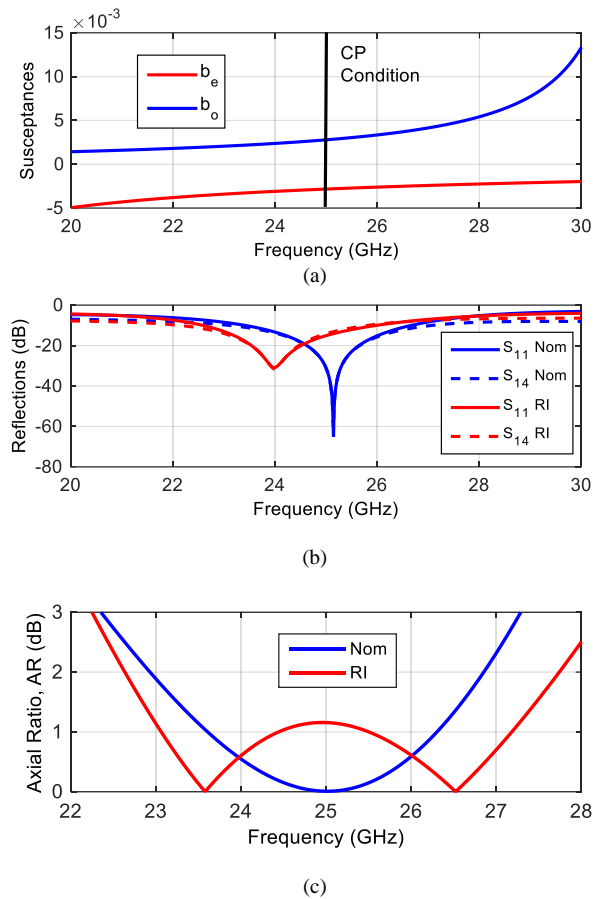

Fig. 11. (a) Susceptances of a set of dimensions for the FSS shown in Fig. 5 when a substrate of $\varepsilon_1 = 2$ is considered at one side of the air-dielectric discontinuity. (b) Reflection coefficients $S_{11}$, $S_{14}$ and (c) Axial Ratio of two LP-CP converters in transmission made up of 2 FSS layers (Fig. 9): (Nom) Design that considers the set of dimensions after a small optimization and (RI) Design that assumes a perturbation of the dimensions to synthesize a ripple. The simulations were obtained with CST at normal incidence in accordance with the nomenclature of the circuit in Fig. 10

branch line of 1 section. The AR is less than 1 dB within the band from 23.6 to 26.3 GHz (10.4%) and the reflection is better than -15 dB. However, it is relatively easy to increase the bandwidth by varying the dimensions slightly and therefore introducing a ripple response, although at the expense of increasing the AR and the reflection at the central frequency (Fig. 11 (RI), L=3.15 mm, W=0.1 mm, S=0.075 mm, P=5 mm, $d = 6.9$ mm). It must be noted that in this case the ripple cannot be associated with a standardized synthesis, since a branch line of N=2 sections does not allow a canonical ripple response.

### B. Example 3

In the case of three stacked FSS and periodical synthesis, the two spacers must have $d_1 = d_2 = \lambda_0/4$ and $\varepsilon_{r1} = \varepsilon_{r2} = 1$, and the reference impedances for the three branches must be (according with the nomenclature of Fig. 10): $Z_{ref1} = Z_{ref3} = Z_0/(\sqrt{2}+1)$ and $Z_{ref2} = Z_0/\sqrt{2}$. The design that considers transmission lines provides a potential AR of less than 1 dB within the band from 21.4 to 28.6 GHz (28.8%), and reflections better than -18 dB. However, to consider the effect of the FSSs, series resonators must be assumed to carry out the branches, as discussed above. In this case, the even and odd inductances and capacitances required to fulfill (15) at each layer at 25 GHz would be those detailed in Table I. Fig. 12a shows the circuital frequency response of the S parameters and the phase-shift between the transmitted components (ports 2 and 3) of the field when a vertically-polarized incident field (port 1) impinges on the designed LP-CP converter. The axial ratio is also represented in Fig. 12b. As can be seen, the AR in this case is less than 1 dB within the band from 19.8 to 26 GHz (26%), which increases with respect to the case of 2 FSS layers (the reflections are better than -15 dB). Note that the 90º of phase-shift between the transmitted field components is maintained within the prescribed bandwidth, as expected from the synthesis process. It can also be noted that the band is displaced as a consequence of the approach considered when the transmission lines are substituted by the series resonators, so that the final process would involve the design at a higher frequency to be made or to optimize the dimensions to center the band.

The design obtained would be physically achievable if the FSS-scatters used were able to provide the values of the even and odd capacitances in Table I. However, it has been proven that the scatter shown in Fig. 5 (which is single resonant), does not reach these values. Thus, the periodic synthesis of 2 sections cannot be achieved with the FSS considered here. It is expectable that a more complex FSS architecture (multi-resonant) exhibits the required values. Furthermore, other synthesis types such as that of Chebyshev could be considered to carry out the design of 3-FSS LP-CP converters in transmission mode. All this work will be evaluated in the future.

## V. EXPERIMENTAL VALIDATION

To validate the proposed design process experimentally and to research the viability of one of the designs presented, a sample of 40x40 cells (200x200 mm) of the reflective LC-CP element designed in section III.A (Fig. 7) was manufactured and tested. The metallizations (FSS) were etched on Arlon 25N (0.5-mm thick) using a laser circuit structuring process, which provided tolerances of ±20 μm to the dimensional values of the cells. The required air thickness (1.5 mm) was achieved using a supporting square frame of FR-4 (1.5-mm thick), which circumscribes the non-active region of the device and was secured to the supporting material and the ground plane (metallic plate) through dielectric screws. Both the tolerances of the FR-4 and the mechanical structure (bending of the supporting material) introduced a tolerance to the air thickness of about ±0.3 mm.

To measure the device at normal incidence, the typical setup made up of both transmitter and receiver horns was considered. However, given our lack of availability of a set of orthomode transducers (OMTs) to cover the broadband of the converter, an offset system was used. To reduce the angle of incidence on the DUT as much as possible, we took advantage of the Compact Antenna Test Range based on the Gregorian system of an anechoic chamber (as shown in Fig. 13a), which generates an incident plane wave that impinges on the converter at an angle of incidence of 0.25°. Fig. 13b shows a photograph of the device placed on the anechoic chamber.

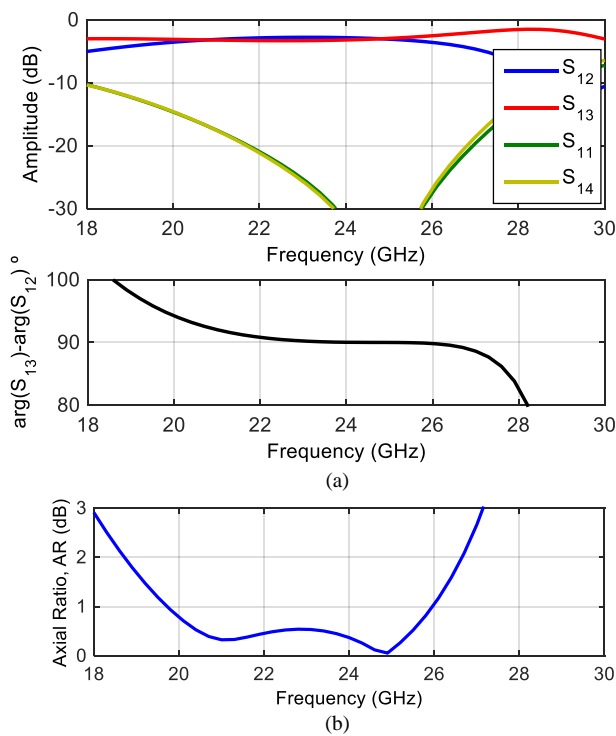

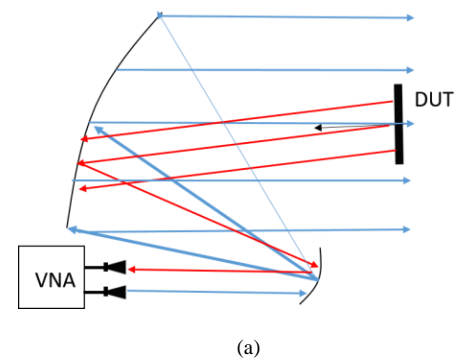

(a)

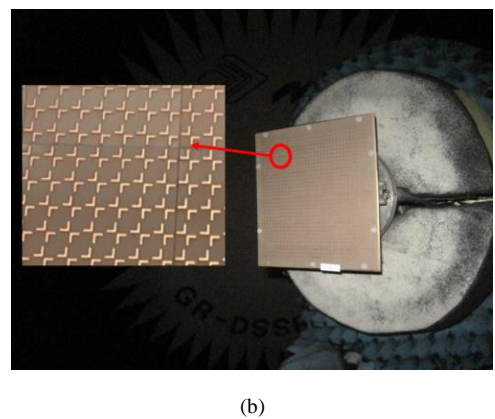

(b)

Fig. 12. (a) S parameters of the 3-FSS layer LP-CP converter designed as a function of the frequency, calculated with the equivalent circuit at normal incidence in accordance with the nomenclature of the circuit in Fig. 10, (b) Axial Ratio.

Fig. 13. (a) Setup used to measure the reflective LP-CP converter at normal incidence, (b) Photograph of the device placed in the anechoic chamber

Fig. 14a shows the measured amplitude and phase difference of the reflection and conversion coefficients ($\rho_{11}$ and $\tau_{12}$ in accordance with the nomenclature used in section III) of the LC-CP converter for a V incident linear polarization. Fig. 14b represents the measured AR (dB). The electromagnetic simulations (CST) are also represented in Fig. 14 for those cases that consider the nominal dimensions (Nom) or the aforementioned tolerances (Nom+Tol). As can be seen, the measured AR is about 1.2 dB in the band from 20 to 30 GHz, and the fixed phase-shift between the two linear reflected components within the band is about 90º, as demonstrated theoretically. The discrepancies between the simulations (Nom+Tol) and measurements in Fig. 14 could be explained as the measurements include the effect of the finite size of the device. Note that although the test system is able to eliminate the effect of the supporting structure of the DUT, the device is illuminated uniformly by a plane wave.

It is clear that a better manufacturing tolerances or a more appropriate selection of materials are required to achieve real AR=0.2 dB, especially those related to the air thickness. It has been computed that the AR of the device increases 0.3 dB when ±20 um of tolerance is considered for the metallizations; the same effect is achieved if 50 um of tolerance is considered for the air thickness. Thus, the sensitivity of the device to dimensional errors is relatively high. Note that the cells exhibiting dual diagonal symmetry tend to show worse sensitivity to dimensional variations than other FSS shapes (i.e. a single dipole), since a variation in one of the main dimension affects the two polarizations at the same time. However, it can be ameliorated if the sensitivity to tolerance errors is considered as a design criterion and therefore included in the design procedure, although probably at the expense of exacerbating the electrical performance in terms of bandwidth.

The difficulty in measuring very low AR would also involve a better measuring setup, which illuminates the device with a certain amplitude taper. However, the test performed here validates the design procedure proposed, and particularly demonstrates the theoretical property of broadband 90º-phase-shift between the reflected linear components of LP-CP converters working in reflection mode (comprising cells with dual diagonal symmetry), which is important in achieving the broadband behavior of these devices.

## VI. CONCLUSIONS

FSS cells that exhibit dual diagonal symmetry can be fully described with a 4-port equivalent circuit at normal incidence. The circuit is suitable if the cells exhibit square period, single mode operation, far-field excitation and enough attenuation of the high-order modes between pairs of stacked FSSs. However, if these limitations are assumed, the circuit allows for the use of standardized coupler theory to design LP-CP converters in both transmission and reflection modes. The standardized synthesis provides some advantages in the design such as simplicity, better performance, and the capability of achieving prescribed electrical conditions systematically for an arbitrary number of layers, although at the expense of using fixed thicknesses (multiple of a quarter of a wavelength) which could increase the profile of the device.

In the case of single layer reflective converters made up of FSS with dual diagonal symmetry, the two reflected field components were theoretically demonstrated to be in quadrature independently of the frequency, so that the bandwidths of these devices are inherently large (BW>30% with AR<0.2 dB).

The synthesis in transmission mode guarantees an exact design (reflection and transmission features) at the central frequency, and identical amplitudes of -3 dB and phase-shift of 90º under a controlled ripple for the two transmitted field components within a prescribed bandwidth and reflection level (pseudo-exact design). The converters operating in transmission usually exhibit narrow bandwidths for a small number of layers, which can be improved by increasing the sections of the device (similarly to what happens to branch line couplers). In the case of two FSS layers made up of single resonant elements, bandwidths of between 10-15% could be achieved if the required AR is assumed to be less than 1 dB. In the case of three stacked FSSs, the periodical synthesis could reach AR<1 dB within bandwidths larger than 25% (reflection levels better than -15 dB), although a more sophisticated FSS-scatter than that used (single resonant) is required. Other FSS shapes (multi-resonant) and/or other synthesis types could improve the performance.

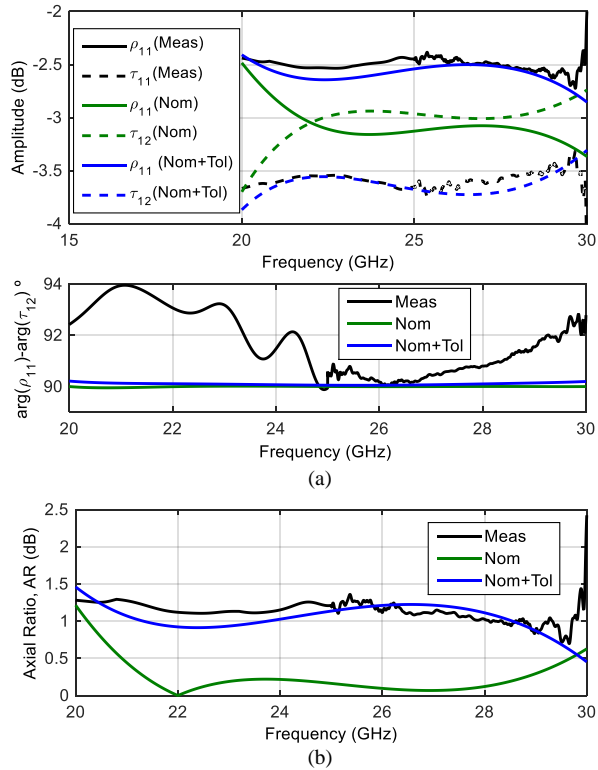

Fig. 14. Measurements and simulations at normal incidence of the reflective LP-CP converter designed in section III.A when a vertically polarized incident field impinges on the device. The simulations consider the nominal dimensions (Nom) and the effects of the tolerances (Nom+Tol). (a) Amplitudes and phase-shift of reflected field components, (b) Axial ratio

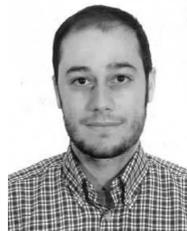

**Gerardo Perez-Palomino** was born in Granada, Spain. He received the M.Sc. and Ph.D. degrees in Telecommunication Engineering from the Universidad Politécnica de Madrid (UPM), Spain, in 2009 and 2015, respectively.

Since 2008, he has been working with the Applied Electromagnetism and Microwaves Group at UPM as a researcher. His research interests include the analysis, characterization, modelling and design of antennas and microwave systems, and the electromagnetic theory. He is involved in the development of reconfigurable antennas in millimeter and sub-millimeter wavelengths and the design of devices based on FSSs, including reflectarrays and transmitarrays.

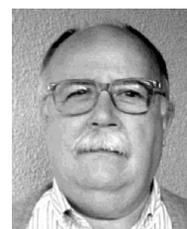

**Juan. E. Page** was born in Madrid, Spain, in 1946. He received the Ingeniero de Telecomunicación and the Ph.D. degrees from the Universidad Politécnica de Madrid, Madrid, Spain, in 1971 and 1974, respectively.

Since 1983, he has been a Full Professor of the Electromagnetism and Circuit Theory Group, currently with the Department of Signals, Systems, and Radio Communications, Universidad Politécnica de Madrid. His research interests include the teaching of electromagnetic and circuit theories and CAD of microwave devices and systems.

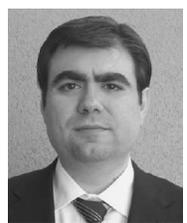

**Manuel Arrebola** (S'99–M'07) was born in Lucena (Córdoba), Spain. He received the Ingeniero de Telecomunicación degree from the Universidad de Málaga, Málaga, Spain, in 2002, and the Ph.D. degree from the Universidad Politécnica de Madrid (UPM), Madrid, Spain, in 2008.

From 2003 to 2007, he was with the Electromagnetism and Circuit Theory Department, UPM, as a Research Assistant. In 2005, he was with the Microwave Techniques Department, Universität Ulm, Ulm, Germany, as a Visiting Scholar. In 2007, he joined the Electrical Engineering Department, Universidad de Oviedo, Gijón, Spain, where he is currently an Associate Professor. In 2009, he was with the European Space Research and Technology Centre, European Space Agency, Noordwijk, The Netherlands. His current research interests include analysis and design techniques of contoured-beam and reconfigurable printed reflectarrays both in single and dual-reflector configurations and planar antennas.

Dr. Arrebola was a co-recipient of the 2007 S. A. Schelkunoff Transactions Prize Paper Award given by the IEEE Antennas and Propagation Society.

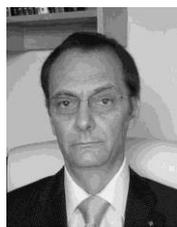

**José. A. Encinar** (S'81–M'86–SM'09–F'10) was born in Madrid, Spain. He received the Electrical Engineer and Ph.D. degrees, both from Universidad Politécnica de Madrid (UPM), in 1979 and 1985, respectively.

Since January 1980, he has been with the Applied Electromagnetics Group at UPM, as a Teaching and Research Assistant from 1980 to 1982, as an Assistant Professor from 1983 to 1986, and as Associate Professor from 1986 to 1991. From February to October of 1987, he was a Postdoctoral Fellow of the NATO Science Program with the Polytechnic University, Brooklyn, NY, USA. Since 1991 he is a Professor of the Electromagnetism and Circuit Theory Group, currently in the Department of Signals, Systems and Radio Communications at UPM. He was a Visiting Professor with the Laboratory of Electromagnetics and Acoustics at Ecole Polytechnique Fédérale de Lausanne (EPFL), Switzerland, in 1996, and with the Institute of Electronics, Communication and Information Technology (ECIT), Queen's University Belfast, U.K., in 2006 and 2011. His research interests include numerical techniques for the analysis of multi-layer periodic structures, design of frequency selective surfaces, printed arrays and reflectarrays.

Prof. Encinar has co-authored more than 200 journal and conference papers, one book and several book chapters. He is holder of five patents on array and reflectarray antennas. He was a co-recipient of the 2005 H. A. Wheeler Applications Prize Paper Award and the 2007 S. A. Schelkunoff Transactions Prize Paper Award, given by IEEE Antennas and Propagation Society. He has been a member of the Technical Programme Comity of several International Conferences (European Conference on Antennas and Propagation, ESA Antenna Workshops, Loughborough Antennas & Propagation Conference).
.